\documentclass[%
twocolumn,
amsmath,amssymb,
]{revtex4-2}
\usepackage{lineno}

\usepackage{rotating}
\usepackage{lipsum}
\usepackage{graphicx}% Include figure files
\usepackage{dcolumn}% Align table columns on decimal point
\usepackage{bm}% bold math
\usepackage{color}
\usepackage{diagbox}
\usepackage[table]{xcolor}
\usepackage{tabularx}
\usepackage{makecell}
\usepackage[none]{hyphenat}
\usepackage{lineno}
\usepackage{mathrsfs}

\definecolor{Navy}{RGB}{54,100,139}
\usepackage[colorlinks,
linkcolor=Navy,
anchorcolor=Navy,
citecolor=Navy]{hyperref}
\usepackage[version=4]{mhchem}
\usepackage{amsmath}
\usepackage{bm}
\usepackage{placeins}
\usepackage[T1]{fontenc}
\usepackage{mathpazo}
\usepackage{textcomp}
\usepackage{nicefrac} 
\usepackage{changes}
\usepackage[misc]{ifsym}
\definechangesauthor[name={HI}, color=orange]{HI}
\usepackage{ragged2e}

\usepackage{xcolor}

\begin{document}
	\setlength {\marginparwidth} {2cm}

	\title{Gapless spin excitations in nanographene-based antiferromagnetic\\
		spin-½  Heisenberg chains
	}

	\author{Chenxiao Zhao$^{1\dagger \ \textrm{\Letter}}$, Lin Yang$^{2,3\dagger}$,  Jo\~{a}o C. G. Henriques$^{4,5\dagger}$, Mar Ferri-Cortés$^{6}$, Gon\c{c}alo Catarina$^{1}$, Carlo A. Pignedoli$^{1}$, Ji Ma $^{2,3}$, Xinliang Feng$^{2,3 \ \textrm{\Letter}}$, Pascal Ruffieux$^{1 \ \textrm{\Letter}}$, Joaquín Fernández-Rossier$^{4 \ \textrm{\Letter}}$, Roman Fasel$^{1,7}$}
	\affiliation{$^{1}$Empa -- Swiss Federal Laboratories for Materials Science and Technology, D\"{u}bendorf, Switzerland.}
	\affiliation{$^{2}$Faculty of Chemistry and Food Chemistry, and Center for Advancing Electronics Dresden, Technical University of Dresden, Dresden, Germany.}
	\affiliation{$^{3}$Max Planck Institute of Microstructure Physics, Halle, Germany.}
	\affiliation{$^{4}$International Iberian Nanotechnology Laboratory, Braga, Portugal.}
	\affiliation{$^{5}$Universidade de Santiago de Compostela, Santiago de Compostela, Spain.}
	\affiliation{$^{6}$ Departamento de F\'isica Aplicada, Universidad de Alicante,  San Vicente del Raspeig, 03690 Alicante, Spain.}
	\affiliation{$^{7}$University of Bern, Bern, Switzerland.}
	
	%}
%\date{\today}% 

\begin{abstract}
	\textbf{
		Haldane's seminal work established two fundamentally different types of excitation spectra for antiferromagnetic Heisenberg quantum spin chains: gapped excitations in integer-spin chains and gapless excitations in half-integer-spin chains.
		In finite-length half-integer spin chains, quantization, however, induces a gap in the excitation spectrum, with the upper bound given by the Lieb-Schulz-Mattis (LSM) theorem.
		%: $\Delta(L) \leq \frac{2JS^2\pi^2}{L}$, where $L$ is the chain length and $J$ is the nearest-neighbor exchange interaction. 
		Here, we investigate the length-dependent excitations in spin-1/2 Heisenberg chains obtained by covalently linking olympicenes --  Olympic rings shaped nanographenes carrying spin-1/2 -- into one-dimensional chains. The large exchange interaction ($J \sim 38$ mV) between olympicenes and the negligible magnetic anisotropy in these nanographenes make them an ideal platform for studying quantum spin excitations, which we directly measure using inelastic electron tunneling spectroscopy. We observe a power-law decay of the lowest excitation energy with increasing chain length $L$, remaining below the LSM boundary. In a long chain with $L=50$,  a nearly V-shaped excitation continuum is observed, reinforcing the system's gapless nature in the thermodynamic limit. Finally, we visualize the standing wave of a single spinon confined in odd-numbered chains using low-bias current maps. Our results provide compelling evidence for the realization of a one-dimensional analog of a gapless spin liquid.
	}
\end{abstract}
\maketitle

%\tableofcontents
%\pagewiselinenumbers
%\linenumbers 

%%%%%%%%%%%%%%%%%%%%%
%%%%%%%%%%%%%%%%%%%%%
%INTRODUCTION
%%%%%%%%%%%%%%%%%%%%%
%%%%%%%%%%%%%%%%%%%%%
\sloppy{}

In one-dimensional (1D) quantum magnets, spontaneous symmetry breaking is inhibited by strong quantum fluctuations \textsuperscript{\cite{pitaevskii91}},  
leading to the emergence of quantum disordered many-body states such as the resonating valence bond states\textsuperscript{\cite{anderson1973,zhou2017quantum}}.
The spin-1/2 Heisenberg chain with antiferromagnetic (AF) coupling stands as a flagship model in quantum magnetism,  described by the Hamiltonian (\ref{eq1}):

\begin{equation}
	\begin{aligned}
		\hat{\mathcal{H}}= J \sum\limits_{i} \hat{\bm{S}}_{i}\cdot\hat{\bm{S}}_{i+1}
		\label{eq1}
	\end{aligned}
\end{equation}
where $J>0$ implies AF nearest neighbour exchange coupling. Although the AF coupling guarantees the formation of singlet pairs, quantum fluctuations render these singlet pairs resonating between different configurations, resulting in a 1D analog of the gapless spin liquid \textsuperscript{\cite{broholm2020quantum}}. In such system, spin correlations in the ground state decay inversely with spin-spin separation, reflecting how quantum fluctuations prevent the formation of N\'eel ordering\textsuperscript{\cite{giamarchi2003quantum}}. 
In the thermodynamic limit, the Bethe ansatz provides analytical expressions for the ground state and excitations\textsuperscript{\cite{bethe1931theorie}}. The ground state is a macroscopic singlet entangling all spins in the chain, and the excited states form a gapless continuum in the excitation spectrum, indicative of bound states comprising at least two fractional spin-1/2 excitations with well defined energy-momentum relation, known as spinons \textsuperscript{\cite{mourigal2013fractional}}. 
For finite-length chains with periodic boundary condition (PBC), the theorem by Lieb, Schultz, and Mattis sets an upper energy bound for the lowest spin excitation energy\textsuperscript{\cite{lieb1961}}:

\begin{equation}
	E_1(L)-E_0(L)\equiv \Delta_{LEE}\leq J \frac{2\pi^2S^2}{L}  
	\label{eq:bound}
\end{equation}
which converges to zero in the thermodynamic limit ($L \to \infty$), where $L$ is the spin chain length. 

Despite the theoretical appeal, the experimental realization of the isotropic spin-1/2 Heisenberg model faces significant challenges. In quasi-1D crystals such as KCuF$_3$ \textsuperscript{\cite{lake2005quantum}}, CuGeO$_3$ \textsuperscript{\cite{hase1993observation}}, CuPzN\textsuperscript{\cite{breunig2017quantum}}, Yb$_2$Pt$_2$Pb \textsuperscript{\cite{kim2013spin}}, and Cs$_4$CuSb$_2$Cl$_{12}$ \textsuperscript{\cite{tran2020spinon}}, interchain coupling induces transitions to dimerized or magnetically ordered phases. Additionally, the lack of access to well-defined finite chains hampers systematic studies of the evolution of spin excitations with chain length as well as of the distinct behaviors of even- and odd- numbered chains.
Recent advances in nanotechnology have led to the creation of artificial spin chains, such as atomic chains on surfaces \textsuperscript{\cite{toskovic2016atomic,choi2019colloquium, wang2024realizing}}, 
quantum dot arrays \textsuperscript{\cite{van2021quantum}}
and optically trapped cold atoms \textsuperscript{\cite{murmann2015antiferromagnetic,jepsen2020spin}}. However, the fabrication of long chains with isotropic exchange interactions exceeding dipolar coupling is challenging in these platforms.
%Therefore, finding a suitable material platform that minimizes these disruptive effects is crucial. 

%%%%%%%%%%%%%%%%%%%%%%%%%%%%%%%%%%%%%%%%%%
\begin{figure}[h]
	\includegraphics[width=8.5 cm]{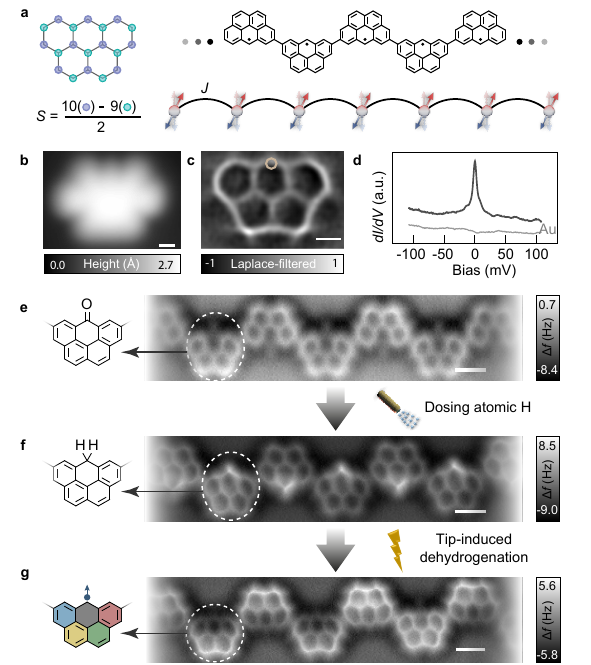}
	\caption{\label{fig1}\textbf{Fabrication of spin-1/2  Heisenberg chain using olympicene.}
		\textbf{a}, Structural illustration of a sublattice-resolved olympicene monomer (left), and chemical structure of the covalently linked olympicene radical chain (right). The associated spin chain model is illustrated below.  \textbf{b} and \textbf{c}, STM image (\textbf{b}, $V_{bias}=-50$ mV, $I_{set}=500$ pA) and Laplace-filtered nc-AFM image (\textbf{c}) of a single olympicene, respectively. White scale bars denote 0.5 nm. \textbf{d}, $dI/dV$ spectra taken on olympicene monomer at the position marked by a brown circle in \textbf{c}, with the background spectrum taken on Au(111) substrate shown by gray curve. Acquisition parameters: $I_{set}=500$ pA and $V_{mod}=1$ mV lock-in root mean squared modulation voltage. \textbf{e}-\textbf{g}, nc-AFM images of the olympicene chain at various stages during synthesis: After polymerization (\textbf{e}), hydrogenation (\textbf{f}), and dehydrogenation (\textbf{g}). Repeat units are marked out by white dotted circles, with corresponding structural models illustrated at the left.
	}
\end{figure}
%%%%%%%%%%%%%%%%%%%%%%%%%%%%%%%%%%%%%%%%%%%%

Recently, $\pi$-electron magnetism in graphene-derived nanomaterials has garnered significant interest due to the substantial and controllable exchange couplings reaching several tens of meV \textsuperscript{\cite{ortiz2019exchange,mishra2020topological,mishra2021large,cheng2022surface,wang2022aza,henriques2023}}.
By covalently assembling selected open-shell nanographenes on an Au(111) surface, the Haldane gap and fractional edge states have been observed in $S=1$ Heisenberg chains\textsuperscript{\cite{mishra2021observation}}. Additionally, a controlled transition between the even- and odd-Haldane phases has been demonstrated in $S=1/2$ dimerized Heisenberg chains\textsuperscript{\cite{zhao2024}}.
Importantly, in nanographene-based systems the covalent connections preserve the many-body interactions of individual magnetic nanographenes, ensuring that the spin excitations remain well-separated from the frontier Hubbard bands of the chain \textsuperscript{\cite{henriques2023}}. Furthermore, the low atomic mass of carbon ensures negligible magnetic anisotropy and spin-orbit coupling \textsuperscript{\cite{lado2014magnetic}}. These attributes make nanographenes an ideal platform for realizing and studying highly entangled quantum spin systems, with potential applications in insulator-based AF spintronics \textsuperscript{\cite{hirobe2017one}}.

Leveraging the potential of on-surface synthesis, we covalently connect specifically designed open-shell nanographenes, named olympicene\textsuperscript{\cite{mistry2015synthesis}}, into chains on a Au(111) surface to establish an isotropic spin-1/2 Heisenberg chain (Fig. \textcolor{Navy}{\ref{fig1}a}). According to the Ovchinnikov-Lieb rule \textsuperscript{\cite{ovchinnikov,lieb1989two,fernandez07}}, each olympicene has a spin$=1/2$ ground state due to a sublattice imbalance of one. 
We use scanning tunneling microscopy (STM) and non-contact atomic force microscopy (nc-AFM) to characterize and manipulate spin degrees of freedom along olympicene chains. Spin excitations are probed using inelastic electron tunneling spectroscopy (IETS)\textsuperscript{\cite{madhavan1998tunneling,ujsaghy2000theory,heinrich2004single,hirjibehedin2006spin,otte2008role,Rossier2009ITS,ternes2015spin}}. Tip-induced dehydrogenation enables precise manipulation of spin sites\textsuperscript{\cite{wang2022aza,Tip-induced}}, facilitating a systematic study of the evolution of spin excitations. 
Additionally, the ability to engineer a chain to have an odd number of units, whose ground state has $S=1/2$,  permits us to confine and probe single spinon \textsuperscript{\cite{kulka23}}. This is in contrast with even-numbered chains where spinons are always created and confined in pairs. 
These advantages enable the experimental observation of: (i) a power-law decay of the lowest spin excitations with chain length $L$, showing a linear dependence on $1/L$ that falls below the LSM bound in the large $L$ region; (ii) nearly "V-shaped" $dI/dV$ spectra for very long chains,  that theory relates to the closing of the gap in the thermodynamic limit; and (iii) standing waves of single spinon in odd-numbered chains. 

%%%%%%%%%%%%%%%%%%%%%
%%%%%%%%%%%%%%%%%%%%%
%END INTRODUCTION, BEGIN EXP.
%%%%%%%%%%%%%%%%%%%%%
%%%%%%%%%%%%%%%%%%%%%

%%%%%%%%%%%%%%%%%%%%%
%%%%%%%%%%%%%%%%%%%%%
%FIGURE 1: 
%%%%%%%%%%%%%%%%%%%%%
%%%%%%%%%%%%%%%%%%%%%
Fig. \textcolor{Navy}{\ref{fig1}b} and \textcolor{Navy}{\ref{fig1}c} show the structural characterization of an individual olympicene molecule on an Au(111) surface using both STM and nc-AFM, respectively. The $dI/dV$ spectrum taken on it (Fig. \textcolor{Navy}{\ref{fig1}c}) displays the sharp Kondo peak, characteristic of a spin-$1/2$ unit on a metallic surface \textsuperscript{\cite{mishra2020topological,Tip-induced}}. Apart from the low-energy spin degrees of freedom, the frontier orbitals of olympicene are also characterized in \textcolor{Navy}{Supplementary I}. 
Covalently connecting olympicenes through their majority carbon sites induces the nearest-neighbor antiferromagnetic exchange coupling $J$, as depicted in Fig. \textcolor{Navy}{\ref{fig1}a}. This was achieved experimentally on Au(111) through thermally induced polymerization of brominated precursor molecules (Fig. \textcolor{Navy}{\ref{fig1}e}), followed by hydrogenation, which replaces the C=O groups with CH$_2$ groups\textsuperscript{\cite{lawrence2022circumventing}} and results in a hydrogen-passivated Olympicene chain (Fig. \textcolor{Navy}{\ref{fig1}f}). In the final step, spin sites were selectively activated by tip-induced dehydrogenation (Fig. \textcolor{Navy}{\ref{fig1}g}). Details of sample preparation can be found in \textcolor{Navy}{Methods} and \textcolor{Navy}{Supplementary II}. 

%%%%%%%%%%%%%%%%%%%%%
%%%%%%%%%%%%%%%%%%%%%
%FIGURE 2: 
%%%%%%%%%%%%%%%%%%%%%
%%%%%%%%%%%%%%%%%%%%%
\begin{figure*}[t]
	\includegraphics[width=16cm]{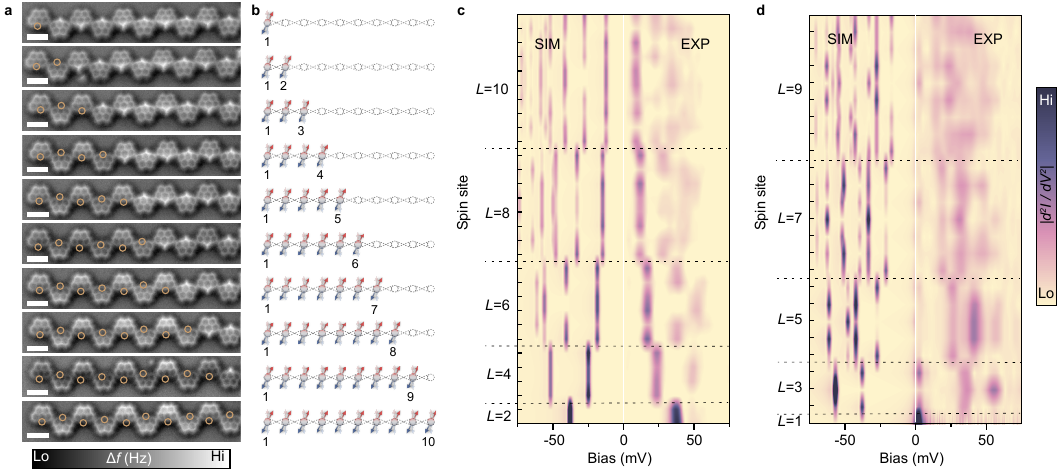}
	\caption{\label{fig2}\textbf{Spin excitations in short chains.}  \textbf{a} and \textbf{b}, nc-AFM images and the corresponding schematic illustrations of selectively activated spin chains, respectively. Effective length $L$ from 1 to 10. Hydrogenated (passivated) olympicene spin sites show bright protrusions in the nc-AFM image, while dehydrogenated (activated) spin sites do not. \textbf{c} and \textbf{d}, Spatially-resolved color map of spin spectral weight ($|d^2I/dV^2|$) obtained at each spin site (brown circles) of chains with even ($L=$2, 4, 6, 8, 10) and odd ($L=$1, 3, 5, 7, 9) lengths. Higher intensities indicate greater probabilities of spin excitation. All spectra are acquired with $I_{\text{set}}= 1$ nA and $V_{\text{mod}}=1$ mV. Simulated spin spectral weights using exact diagonalization (ED) calculations are shown on the negative bias side. White scale bars denote 1 nm.}
\end{figure*}

Figure \textcolor{Navy}{\ref{fig2}a} shows the step-by-step activation of a pre-passivated olympicene chain, through which we obtained spin chains of different effective lengths from $L=1$ to 10 (Fig. \textcolor{Navy}{\ref{fig2}b}). 
Firstly, the nearest neighbor exchange coupling $J$ can be measured in a chain with an effective length of $L = 2$. The spin spectral weight, proportional to $d^2I/dV^2$ in experiments, reflects the energies of excited states and their corresponding excitation probabilities. For $L=2$, a single excitation occurs from the singlet ground state to the triplet excited state, with the energy difference indicating an exchange interaction of $J \approx 38$ meV (Fig. \textcolor{Navy}{\ref{fig2}c}). This value is consistent with our multi-reference Hubbard model calculations for olympicene dimers (see \textcolor{Navy}{Supplementary III}). In addition, we have verified, both experimentally and theoretically, that next nearest neighbor coupling is negligible in our system ( \textcolor{Navy}{Supplementary IV}).

Next we investigate the length-dependence of the low-lying spin excitations. The spin spectral weight for each spin site in different chains is shown in Fig. \textcolor{Navy}{\ref{fig2}c} and \textcolor{Navy}{\ref{fig2}d}, which display a distinct odd-even behavior around zero bias: chains with an even number of units consistently show excitation gaps, with the gap size decreasing as $L$ increases, while chains with an odd number of units display a zero-bias peak whose intensity decreases with increasing $L$. This is consistent with the fact that the ground state is a $S=0$ singlet ($S=1/2$ doublet) for even (odd) numbered chains. The experimental observations align with the simulated spectral weight based on exact diagonalization (ED) of Hamiltonian (\ref{eq1}) with the experimentally determined $J$ (negative bias sides in Fig. \textcolor{Navy}{\ref{fig2}b} and \textcolor{Navy}{\ref{fig2}c}). We find that the excitations are associated with $S=1$ ($S= 1/2$ and $S= 3/2$) states for even- (odd)-numbered chains.  Both theory calculation and experiment results show that, for most excitations, the intensity of the $dI/dV$ steps is modulated along the chain, reflecting the loss of translational invariance and the formation of standing waves. This modulation is missing in the case of rings,  i.e. chains with PBC. 

% \bluemark{Optional: For even-numbered chains the modulation of the lowest energy step is faint, consistent with the fact that the state has a strong overlap with the $q=\pi$ excitation of the chain with  PBC. In contrast, higher energy excitations are linear combination of several  pairs of PBC states with opposite momenta, which naturally accounts for their stronger modulation. (Joao: if this stays, I think we need to show some calculation for it, like the one I showed in todays meeting).
	
	%\redmark{The odd-even behavior can be understood in terms of interacting spinon picture\textsuperscript{\cite{karbach2000introduction}}. 
		%The quantized momentum in limited chains are given by $q=pi n/L$ (n=1, 2,..L). But the spinon dispersion exist only within half of the Brillouin zone [0,$\pi$], so the lowest energy for even-numbered chains related to the spinon modes at $k=\pi/2$.  
		%Chains with an even number of units have a ground state with $S=0$. Excitation to the lowest excited states ($S=1$, due to selection rules\cite{Rossier2009ITS}) corresponds to the creation of two spinons with minimal interaction. As the chain length increases, these two spinons can be separated further apart, reducing their repulsive energy between each other and thereby decreasing the lowest excitation gap. Excited states with stronger spinon-spinon interaction contribute to the multi-step features above the lowest excitation gap, as shown in the $d^2I/dV^2$ spectra when $L>2$.

		%%%%%%%%%%%%%%%%%%%%%
		%%%%%%%%%%%%%%%%%%%%%
		%FIGURE 3: 
		%%%%%%%%%%%%%%%%%%%%%
		%%%%%%%%%%%%%%%%%%%%%
		\begin{figure}
			\includegraphics[width=8cm]{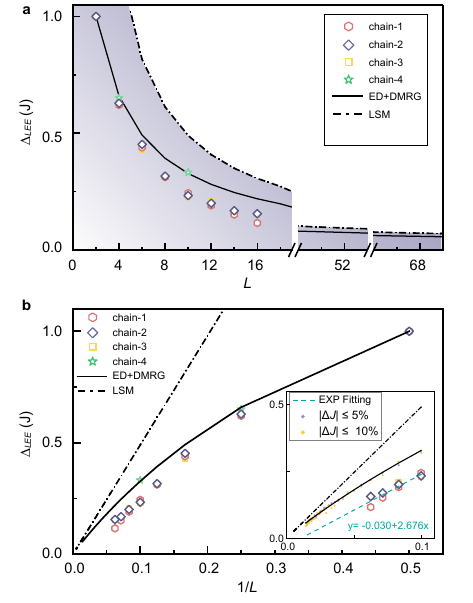}
			\caption{\label{fig3}\textbf{ Evolution of the $\Delta_{LEE}$ with $L$.}
				\textbf{a,} Evolution of the lowest excitation energy ($\Delta_{LEE}$) with chain length $L$ (even-numbered chains only). Data from four different chains are shown (details in Supplementary \textcolor{Navy}{V}). Results from exact diagonalization (ED) and the theoretical value of the Lieb-Schultz-Mattis (LSM) bound are also shown as references. The region below the LSM bound is shaded in purple. \textbf{b}, Relationship between the $\Delta_{LEE}$ and $1/L$ (even-numbered chain lengths only). Inset: Magnified view of the large $L$ region ($1/L \leq 0.1$), showing only experimental data from chains 1-3. A linear fit is applied, with the corresponding function displayed alongside the fitting line. Different random perturbations in $J$ ($\Delta J$ within $\pm 5\%$ and $\pm 10\%$) are considered in DMRG calculations (details in Methods).
			}
		\end{figure}
		
		%\redmark{As a manifestation of the spinon-spinon interaction along the chain, the lowest excitation energies ($\Delta_{LEE}$s) in even-numbered chains qualititively reflects the spin-spin correlation over a distance $\sim L$. }
		The evolution of the lowest excitation energies ($\Delta_{LEE}$) with $L$ are monitored in four different chains, as shown in Fig. \textcolor{Navy}{\ref{fig3}a} (chains 1-4). The $\Delta_{LEE}$ clearly decay with $L$ in a power-law manner, in contrast to the exponential decay in gapped spin chains\textsuperscript{\cite{mishra2021observation,zhao2024}}. This is also confirmed by calculations obtained by using ED and density matrix renormalization group (DMRG) methods (solid black curve in Fig. \textcolor{Navy}{\ref{fig3}a}). 
		%The power-law decay of LEEs suggests the presence of quasi-long-range spin-spin correlations in our system, a characteristic feature of the gapless 1D spin liquid phase\textsuperscript{\cite{luther1975calculation}}. 
		%\redmark{The experimental decay is slightly faster than the calculation owing to the influence from the Au(111) substrate.} 
		We note that experimental data lies below the prediction of theory, which we attribute to the renormalization of the excitation energy due to Kondo exchange with the substrate,  as observed in magnetic adatoms \textsuperscript{\cite{oberg2014}} and nanographene dimers \textsuperscript{\cite{jacob22,krane2023exchange}}
		%
		%In addition, all LEEs lie below the LSM bound (Fig.  \textcolor{Navy}{\ref{fig3}a}), indicating the gapless nature of the system in the thermodynamic limit. 
		%Considering the sine-type dispersion of the two-spinon continuum in energy-momentum space, the first-order expansion in the small $k$ region is expected to show a linear dependence on $1/L$,  where $k$ is quantized as $2n\pi/L$ in finite chains with $n = [1, L]$.
		%
		%\redmark{Considering the scattering and collision rate between two spinons is inversely proportional to $L$, the spinon-sponin interaction induced gap should decay linearly with $1/L$ close to the thermal dynamic limit. } 
		Nonetheless, the observed chain length dependence of $\Delta_{LEE}$ is consistent with the LSM theorem\textsuperscript{\cite{lieb1961}}. In the large $L$ region, $\Delta_{LEE}$ fits well with a linear relationship to $1/L$ (green dashed line in the inset of Fig. \textcolor{Navy}{\ref{fig3}b}), lying below the LSM boundary. Although the LSM theorem is formulated for chains with PBC, we demonstrate that $\Delta_{LEE}$ for open boundary condition (OBC) chains of length $L$ is always smaller than for PBC chains of the same length (\textcolor{Navy}{Supplementary VI}).
		%In general, the LEEs exhibit a slower decay in the shorter chain region ($1/L > 0.2$) and accelerate as the chains lengthen ($1/L < 0.1$), reflected by a noticeable change in slope. 
		%and find that the experimental data is well fitted by a linear relationship, as indicated by theof Fig. \textcolor{Navy}{\ref{fig3}b}. 
		%The near-parallel alignment between the experimental fitting and the ED+DMRG calculation curve suggests a comparable decay rate as the system approaches the thermodynamic limit. 
		%The extrapolation of the fitted line intersects the $x$-axis at $1/L \sim 0.011 \pm$, corresponding to a chain length of $L \sim 90$, where the $\Delta_{LEE}$  is expected to vanish in the olympicene chain system.
		
		Furthermore, we have verified that the evolution of $\Delta_{LEE}$ towards zero is robust against perturbations in $J$. The color-coded points around the DMRG curve in the inset of Fig. \textcolor{Navy}{\ref{fig3}b} show that random fluctuations in $J$ of $\pm 5\%$ and $\pm 10\%$ have a negligible effect on the decay rate, thus preserving the gapless character in the thermodynamic limit. In experiments, perturbations in $J$ (less than $\pm10\%$) can arise from slight structural chain distortions induced by the Au(111) herringbone surface reconstruction and the mixing of inversion- and mirror-type connections along the chain (see \textcolor{Navy}{Supplementary VII}), which accounts for the slight variations observed between chains 1-4.
		
		%%%%%%%%%%%%%%%%%%%%%=
		%%%%%%%%%%%%%%%%%%%%%
		%FIGURE 4: 
		%%%%%%%%%%%%%%%%%%%%%
		%%%%%%%%%%%%%%%%%%%%%
		\begin{figure}
			\includegraphics[width=8 cm]{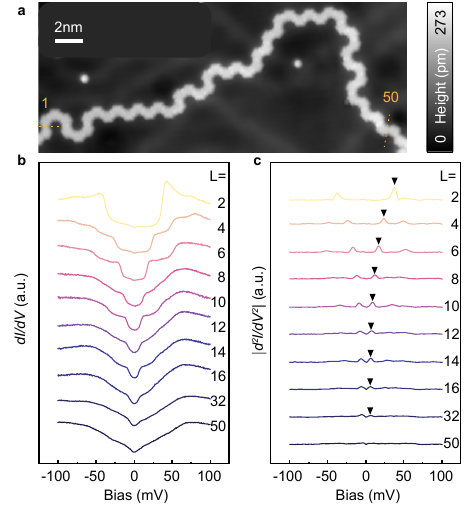}
			\caption{\label{fig4}\textbf{Spin excitation approaching the thermodynamic limit.}
				\textbf{a}, STM image of an olympicene chain with an effective length of $L=50$. $V_{bias} = -300$ mV, $I_{set} = 100$ pA. Yellow dotted lines indicate the effective ends which are the boundaries between activated and passivated spin sites. \textbf{b}, IETS spectra taken from the middle region (averaged over sites $L/2$ and $L/2+1$) of chains with varying lengths. \textbf{c}, $|d^2I/dV^2|$ for chains of different lengths, obtained by numerically differentiating the spectra shown in \textbf{b}. The lowest excitations are indicated by black triangles. All spectra are taken with $V_{mod}=1$ mV, $I_{set}=1$ nA.
			}
		\end{figure}
		
		For long chains where the $\Delta_{LEE}$ decay to a value comparable to the experimental energy resolution ($\sim$ 2 meV), the excitation spectra show a behavior consistent with the thermodynamic limit. We examined the spin excitation spectra in a chain with an effective length of
		$L = 50$ (Fig. \textcolor{Navy}{\ref{fig4}a}), where we expect the spin excitation gap to be $\sim$ 2 meV (based on Fig. \textcolor{Navy}{\ref{fig3}a}). In Fig. \textcolor{Navy}{\ref{fig4}b}, we compare spin excitations measured at the center of chains (averaged over sites $L/2$ and $L/2 + 1$) with varying lengths. As $L$ increases, the number of discernible excitation steps grows, evolving from single steps to multiple steps, and eventually to featureless slopes. For chains with lengths ranging from $L=12$ to $L=32$, only the lowest excitation steps are resolved, with their heights decreasing as $L$ increases. All densely packed higher excitation steps merge into a continuous slope. For $L=50$, individual steps are no longer distinguishable, and the entire spectrum transforms into a "V-shaped" continuum. This evolution can also be monitored from the spin spectral weight ($|d^2I/dV^2|$), where the excitations are indicated by peaks (Fig. \textcolor{Navy}{\ref{fig4}c}). 
		The lowest energy excitations decrease in both energy and intensity as $L$ increases. For $L=50$, no distinct excitation peaks are observed, highlighting the near-continuum character of the spectrum in the long chain limit. 
		%\bluemark{Optional: Modelling such long chains is out of reach of state of the art computation. We have however found a similar evolution of the $dI/dV$ spectra calculated using a semiclassical  model for ferromagnetic chains, that can be easily tackled numerically }
		
		%%%%%%%%%%%%%%%%%%%%%=
		%%%%%%%%%%%%%%%%%%%%%
		%FIGURE 5: 
		%%%%%%%%%%%%%%%%%%%%%
		%%%%%%%%%%%%%%%%%%%%%
		\begin{figure}
			\includegraphics[width=8.5 cm]{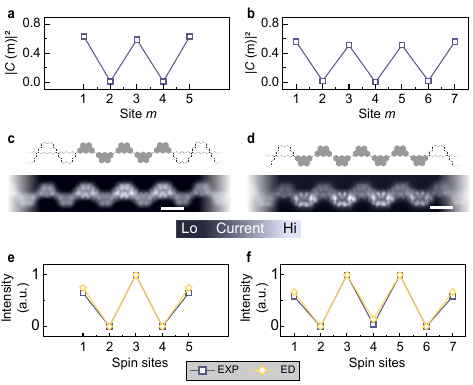} %Figure concept
			\caption{\label{fig5}\textbf{Imaging standing wave of a single spinon.} %and an interacting spinon pair.}
			\textbf{a} and \textbf{b}, Theoretical calculation of $|C(m)|^2$, denoting the overlap of the ground state of the chains with $L=5$ and $L=7$ with localized spinons at site $m$. 
			\textbf{c} and \textbf{d}, Schematic structures and low-energy (-5 meV) constant-height current images for olympicene chains with 5 and 7 effective units, respectively. The activated/passivated spin sites are denoted by gray/blank olympicene geometries. \textbf{e} and \textbf{f}, Axial intensity profiles extracted from the current images in \textbf{c} and \textbf{d}, respectively, symmetrized along the chain and normalized between 0 and 1. The zero-bias $dI/dV$, obtained from  ED-calculations, are shown for comparison. %\textbf{f}, STM image of a even-numbered chain with $L=18$. $V_{bias} = -50$ mV, $I_{set} = 500$ pA. \textbf{g}, Spatially resolved $dI/dV$ map, taken with $V_{bias} = -5$ mV, $I_{set} = 200$ pA, $V_{mod} = 500$ pV. Both 2D and 3D view are shwon.
		}
	\end{figure}
	
	Finally, we focus on the odd-membered chains which possess an $S=1/2$ ground state, and demonstrate that it is a manifestation of a confined single spinon.
	As indicated by recent work of Kulka et al. \textsuperscript{\cite{kulka23}}, the energy, momentum $q$ and spin $S=1/2$ of spinons in PBC chains with length $L=2n$ ($n=1,2,3...$) are well described through the construction of a linear combination of states where an additional spin $\uparrow$ is inserted at site $m$ ($m = 1,...,2n + 1$) in the ground state of the spin chain (a similar reasoning can be applied when a spin $\downarrow$ is inserted instead). We label these states as $|\Psi(m)\rangle$ and refer to them as {\em localized spinon} states (see \textcolor{Navy}{Supplementary VIII}). Intriguingly,  
	the state $|\Psi(q)\rangle\propto\sum_m e^{i  q m} |\Psi(m)\rangle$, defined in  the $L=2n+1$ PBC chain,  has the same wave vector $q$, excitation energy and spin than a spinon in the $L=2n$ PBC chains. 
	Importantly, we have found that the ground state of the $L=2n+1$ OBC chain, $|\Psi_0\rangle$, can be expressed as a linear combination of the localized spinon states,  $|\Psi_0\rangle=\sum_{m=1,2n+1} C(m) |\Psi(m)\rangle$. 
	In contrast to the case of extended single spinon states $|\Psi(q)\rangle$ in PBC chains, the coefficients $C(m)$ peak in the odd $m$ sites along the OBC chains (see Fig.  \textcolor{Navy}{\ref{fig5}a} and  \textcolor{Navy}{\ref{fig5}b}). In other words,  the ground state of the $L=2n+1$ OBC chain can be interpreted as a single-spinon standing wave (see \textcolor{Navy}{Supplementary VIII}) that has a strong modulation in the local magnetization $\langle \Psi_0|S_z(m)|\Psi_0\rangle$, which also peaks in the odd-sites. 
	Therefore, the single spinon standing wave should result in a modulation of the amplitude of the Kondo peak along the chain. 
	In the conventional theory for IETS\textsuperscript{\cite{Rossier2009ITS,ternes2015spin}},
	the zero bias conductance is related
	%we can relate the height of the  zero bias $dI/dV$ 
	to the square of the local magnetization of the ground state,  $|\langle \Psi_0|S_z(m)|\Psi_0\rangle|^2$, through resonant tunneling via the Kondo effect. 
	
	In order to test this prediction, we have carried out low-bias current mapping in odd-numbered chains.  As shown in Figs. \textcolor{Navy}{\ref{fig5}c} and \textcolor{Navy}{\ref{fig5}d}, the low-bias ($-5$ mV) current maps for $L=5$ and $L=7$ chains reveal strong modulation along the chains, with pronounced intensity at the odd-numbered sites. Summing the intensity corresponding to each spin site provides a clearer view of the modulation along the chain axis (Figs. \textcolor{Navy}{\ref{fig5}e} and \textcolor{Navy}{\ref{fig5}f}), which exhibits a periodicity of two. This result is well reproduced by the simulated Kondo peak intensity from ED calculations and consistent with the single spinon standing wave picture outlined above (Figs. \textcolor{Navy}{\ref{fig5}a} and \textcolor{Navy}{\ref{fig5}b}).

	We have successfully fabricated and characterized length-controlled nanographene chains realizing the gapless antiferromagnetic spin-1/2 Heisenberg quantum spin model. Our inelastic electron tunneling spectroscopy results have revealed lowest energy spin excitations that decay linearly with $1/L$ and a "V-shaped" excitation continuum for long chains, evidencing a vanishing spin excitation gap in the thermodynamic limit, in line with quasi-long-range spin-spin correlations that decay in a power law rate. Importantly, the capability of building odd-numbered spin chains has allowed us to image a single spinon standing wave. Given that spinons are rather exotic quasiparticles that have largely evaded an intuitive picture, this is an important starting point for investigating spinon properties and interactions in prototypical quantum spin chains.

	\FloatBarrier

	\textbf{Acknowledgements}
	This work was supported by the Swiss National Science Foundation (grant numbers 212875, 205987), the NCCR MARVEL, a National Centre of Competence in Research, funded by the Swiss National Science Foundation (grant number 205602), the Graphene Flagship Core 3 (grant no. 881603), ERC Consolidator grant (T2DCP, grant no. 819698), the  Center for  Advancing  Electronics  Dresden  (cfaed),  H2020-EU.1.2.2.-FET  Proactive  Grant  (LIGHT-CAP,  101017821), the   DFG-SNSF   Joint   Switzerland-German   Research   Project (EnhanTopo, No. 429265950), the European Union (Grant FUNLAYERS-101079184), the Funda\c{c}\~{a}o para a Ci\^{e}ncia e a Tecnologia (Grant No. PTDC/FIS-MAC/2045/2021), Generalitat Valenciana  (Grants No. Prometeo2021/017, No. MFA/2022/045 and CIACIF/2021/434), MICIN-Spain (PID2022-141712NB-C22 ) and the Advanced Materials programme  supported by MCIN with funding from European Union NextGenerationEU (PRTR-C17.I1). We also greatly appreciate financial support from the Werner Siemens Foundation (CarboQuant). For the purpose of Open Access (which is required by our funding agencies), the authors have applied a CC BY public copyright license to any Author Accepted Manuscript version arising from this submission.
	\nolinenumbers
	
	\textbf{Author contributions}
	\justifying
	\noindent
	R.F., P.R., C.Z., L.Y., and X.F. proposed the project. L.Y. synthesized and characterized the precursor molecules in solution, with supervision provided by J.M. and X.F. C.Z. performed the on-surface synthesis, STM/STS measurements and analyzed the data. J.C.G.H., M.F.C., and G.C. performed the CAS, ED, and DMRG calculations with supervision provided by J.F.R. C.Z and C.A.P. performed the DFT calculation. J.C.G.H., G.C., J.F.R analyzed the spinon theory. C.Z., J.F.R., and R.F. wrote the Manuscript. All authors discussed the results.
	\\
	
	\noindent
	$^\dagger$ equal contribution \\
	\email[{$\textcolor{Navy}{^\textrm{\Letter}}$} Corresponding author:  {
		\\chenxiao.zhao@empa.ch\\
		xinliang.feng@tu-dresden.de\\
		pascal.ruffieux@empa.ch\\
		jfrossier@gmail.com
	}\\
	
	%\nocite{*}

	\bibliography{Heisenberg}% Produces the bibliography via BibTeX.
	%%%%%%%%%%%%%%%%%%%%%
	%%%%%%%%%%%%%%%%%%%%%
	%Supplementary
	%%%%%%%%%%%%%%%%%%%%%
	%%%%%%%%%%%%%%
	\centering{\section{Methods}}
	
	\justifying
	
	\textbf{Sample preparation} Au(111) single-crystal surfaces were prepared by Ar$^+$-ion sputtering followed by annealing at 430 $^{\circ}$C. Precursor molecules (see Supplementary II for details) of the olympicene were deposited on the clean Au(111) surface, held at room temperature, via sublimation from a quartz crucible at $\sim$ 150 $^\circ$C. After molecule deposition, the sample was annealed at  $\sim$ 250 $^\circ$C for 5 min to induce surface-assisted polymerization. Then the sample was flashed to 300 $^\circ$C several times to remove the detached Br atoms from the surface. Hydrogen reduction leading to deoxygenation of the ketone groups was achieved by exposing the sample to atomic H produced by passing 3$\times$ 10$^{-8}$ mbar high-purity (99.999\%) H$_2$  gas for 10 hours through a thermal gas cracker held at 1500-1600$^\circ$C.

	\textbf{Spin manipulation methods} Tip-induced dehydrogenation is employed to manipulate the spin sites. Initially, all spins in the olympicene chain are passivated following hydrogenation, due to the formation of CH$_2$ groups that pair with the originally unpaired $\pi$ electrons. By applying a bias voltage of approximately 2.5 V using an STM tip, a single hydrogen atom can be selectively removed from the CH$_2$ group, transforming the sp$^3$-configured CH$_2$ group into an sp$^2$-configured CH group, thus contributing an unpaired $\pi$ electron and activating the spin. Repeating this activation process on neighboring molecules allows for the creation of spin chains with perfectly controlled chain lengths.
	
	\textbf{STM and STS measurements} STM and STS measurements were performed in a commercial low-temperature STM/nc-AFM system from Scienta Omicron operating at a temperature of 4.5 K and a base pressure below 2$\times$ 10$^{-11}$ mbar. All $dI/dV$ spectra and bond-resolved current images were acquired with a CO-functionalized tungsten tip. \textit{In-situ} cold deposition of CO molecules was performed to obtain a CO-functionalized tip. Differential conductance $dI/dV$ spectra were acquired with a lock-in modulation frequency of 691 Hz, with an amplitude that is specified in the corresponding figure captions. The nc-AFM images were taken with a qPlus tuning fork sensor\textsuperscript{\cite{giessibl2019qplus}} (resonance frequency $\sim$27 kHz, quality factor $\sim$27K) with a CO-functionalized tungsten tip in constant-height mode.

	\textbf{CAS calculations}
	The starting point for the CAS calculations is a single-orbital tight-binding model that focuses exclusively on the $p_z$ orbitals of the carbon atoms in the nanographenes. In our approximation, we consider both first- and third-nearest neighbor hoppings, denoted as $t_1$ and $t_3$, respectively. The corresponding tight-binding Hamiltonian is given by:
	\begin{align}
		H_{0} = - t_1 \sum_\sigma \sum_{\langle i,j \rangle}  c^\dagger_{i, \sigma} c_{j, \sigma} - t_3 \sum_\sigma \sum_{\langle\langle\langle i,j \rangle\rangle\rangle}  c^\dagger_{i, \sigma} c_{j, \sigma},
	\end{align}
	where $c^{(\dagger)}_{i,\sigma}$ denotes the annihilation (creation) operator of an electron with spin projection $\sigma = \uparrow, \downarrow$ at site $i$. This single-particle model is diagonalized, resulting in a set of molecular orbitals.
	
	At charge neutrality, an olympicene chain with length $L$  features $L$ half-filled zero-energy states, which are slightly hybridized due to $t_3$. For the CAS calculations, we consider a subset of molecular orbitals that includes these $2N$ zero-energy states, along with the two closest states in energy, to account for the Coulomb-driven exchange mechanism\textsuperscript{\cite{jacob22}}. Thus, the active space comprises $N_{MO} = L + 1$ molecular orbitals.
	Then, we include interactions within the Hubbard model approximation, considering an on-site Hubbard repulsion $U$:
	\begin{equation}
		H_{U} = U \sum_i n_{i,\uparrow} n_{i,\downarrow},
	\end{equation} 
	with $n_{i,\sigma} = c^\dagger_{i,\sigma} c_{i,\sigma}$.
	The many-body Hubbard Hamiltonian $H_0 + H_U$ is represented in the restricted basis set, considering all the multi-electronic configurations that can be obtained with $N_e$ electrons in the $N_{MO}$ molecular orbitals.
	Assuming half-filling, we always have $N_e=N_{MO}$.
	The remaining electrons are thus assumed to fully occupy the molecular orbitals below the active space, implying that these are frozen doubly-occupied orbitals; the occupation of the molecular orbitals above the active space is also assumed to be frozen, featuring zero electrons.
	Finally, the resulting (truncated) Hubbard Hamiltonian is diagonalized numerically.
	
	\textbf{DMRG calculations}
	DMRG calculations were carried out using the ITensor Julia library\textsuperscript{\cite{fishman_itensor_2022}}.
	We employed a protocol where the maximum bond dimension is allowed to grow indefinitely as to keep the truncation error cutoff below $1 \times 10^{-8}$, which ensures very high accuracy.
	We started the variational optimization with a randomized matrix product state and stopped it when energy and total spin were converged up to $1 \times 10^{-4}$~meV and $1 \times 10^{-5}$, respectively.
	
	\textbf{Modelling $dI/dV$ spectroscopy}
	To model the $dI/dV$ spectroscopy we consider a spin chain that is coupled to two electron reservoirs: the STM tip and the substrate. In our description, we consider two types of electron scattering that can produce a spin flip in a given site of the chain\textsuperscript{\cite{Appelbaum1967Exchange}}: electrons that tunnel from tip to sample, exciting the spin chain in the process; and scattering between the substrate electrons. To compute the current, $I$, we apply scattering theory including corrections up to third order\textsuperscript{\cite{ternes2015spin}}; then we differentiate the current with respect to the bias $V$ (defined as the difference between the chemical potential of the two reservoirs), thus obtaining the theoretical prediction for $dI/dV$. Up to second order, the $dI/dV$ spectrum is composed of thermally broadened excitation steps, whose height is determined by the spin spectral weight\textsuperscript{\cite{Rossier2009ITS}}.
	The spin spectral weight corresponding to a transition from a state $M$ to a state $M'$ on a given spin site $i$ is given by
	\begin{equation}
		\mathcal{W}_{M,M'}(i) = p_M \sum_{a=x,y,z} | \langle M'| \hat{S}_a(i) | M\rangle |^2,
	\end{equation}
	where $p_M$ denotes the population of state $M$.
	This quantity reflects the probability of exciting state $M'$ (from state $M$, which is typically the ground state) with a local spin-flip interaction promoted by an STM tip placed on site $i$, with bias matching or exceeding the energy difference between $M$ and $M'$. 
	Excitations are only possible if the total spin, $S$, of the involved states respects the relation $\Delta S=0,\pm1$. The third-order correction accounts for processes mediated by intermediate states; the spin conservation rule is enforced between the initial and final states, as well as between the initial/final state and the intermediate states; energy conservation, however, is only required between the initial and final states. The third order correction is responsible for the introduction of logarithmic resonances which lead to two main changes in the spectrum: i) the thermally broadened steps acquire an overshooting feature at the onset of excitation; ii) a Kondo peak appears at zero bias if the system has a degenerate ground state. In our calculations, we have considered a temperature of 4.5 K.
	
	The ED calculations were carried out using the QuSpin package\textsuperscript{\cite{weinberg_quspin_2017}}. 
	
	\textbf{Modelling influence of fluctuations in exchange } 
	For the simulation of possible perturbations in the exchange coupling $J$ along the olympicene chain, we added  to the reference value $J$ a random term $\delta J_{i,i+1}$ to the exchange between sites $i$ and $i+1$. The values of $\delta J$ were drawn from a uniform distribution over an interval of $\pm 5\%$ $J$ and $\pm 10\%$ $J$.  For every chain length we computed the gap by averaging over at least 20 realizations of the disorder configuration. 
	
	%%%%%%%

\end{document}